\documentclass[prl,floatfix,twocolumn,showpacs]{revtex4}
\usepackage{amsmath}
\usepackage{amsfonts}
\usepackage{mathrsfs}
\usepackage{epsfig}

\usepackage{graphicx}
\usepackage{dcolumn}
\usepackage{amsmath}

\def\be{\begin{equation}}
\def\ee{\end{equation}}

\begin{document}
\title{Chimera order in spin systems}
\author{Rajeev Singh$^1$, Subinay Dasgupta$^2$ and Sitabhra Sinha$^1$}
\affiliation{$^1$The Institute of Mathematical Sciences, CIT Campus,
Taramani,
Chennai 600113, India.\\
$^2$Department of Physics, University of Calcutta, 92 Acharya Prafulla
Chandra Road, Kolkata 700009, India.}
\date{\today}
\begin{abstract}
Homogeneous populations of oscillators have recently been shown to
exhibit stable coexistence of coherent and incoherent regions.
Generalizing the concept of chimera states to the context of
order-disorder transition in systems at thermal equilibrium, we show
analytically that such complex ordering can appear in a system of
Ising
spins, possibly the simplest physical system exhibiting this
phenomenon.
We also show numerically the existence of chimera ordering in
3-dimensional spin systems that model layered magnetic materials,
suggesting possible means of experimentally observing such states.
\end{abstract}

\maketitle

\newpage

Transition to states characterized by simple or complex ordered
patterns
is a phenomenon of central importance
in equilibrium statistical physics as well as in dynamical systems far from
equilibrium~\cite{Stanley99,Acebron05}.
Examples of simple ordering at thermal equilibrium include
the aligned orientation of spins in Ising-like systems, while, in the
context of nonlinear dynamics, this may be observed in the phase
synchronization of coupled oscillators.
However, more complex ordering behavior may also occur in various
systems under different conditions, especially in the presence of
heterogeneities.
A surprising recent finding is that even {\em homogeneous} dynamical
systems can exhibit a robust, partially ordered state characterized 
by the coexistence of 
incoherent, desynchronized domains with coherent, phase locked
domains~\cite{Kuramoto}. 
Such {\em chimera} states have so far been observed only in
different types of oscillator populations, including complex
Ginzburg-Landau equations, phase oscillators, relaxation oscillators,
etc., arranged in various connection
topologies~\cite{Strog04,Sethia08,Omelchenko08,Strog08}.
Given that ``chimera" refers to the co-occurrence of incongruous
elements, one can generalize the concept of chimera-like
states to include those characterized by simultaneous existence of 
ordered and disordered regions in an otherwise homogeneous system. 
If such a state can occur as the global energy minimum of a
system in thermal equilibrium, it may widen the scope of experimentally
observing chimera-like order in physical situations. 

It is with this aim in mind that we investigate chimera-like ordering
in systems at thermal equilibrium. Specifically, we consider spin-models 
as they are
paradigmatic for different complex systems 
comprising interacting components which can be in any of multiple
discrete states. For example, simple Ising-like models consisting
of binary-state elements are versatile enough to be used for
understanding processes operating in a wide range of physical (e.g., magnetic
materials~\cite{dejongh74,kincaid75,stryjewski77}), biological (e.g., neural
networks~\cite{amit89}) and
social (e.g., opinion formation~\cite{castellano09,pan09}) systems. 
The nature and connection topology of the interactions between the spins 
decide whether the entire population reaches a consensus,
corresponding to an ordered state, or is in a disordered state that 
corresponds to the stable coexistence of contrary orientations. 
The existence of a chimera state in such situations would imply that 
even though every spin is in an identical environment, different regions 
of the system will exhibit widely different degrees of ordering.

In this paper we report for the first time the occurrence of chimera
order in spin systems. This is characterized for a system of Ising
spins by the occurrence of regions having different ordering behavior 
as measured by the magnitude of magnetization. The specific system we
consider in detail is globally coupled and comprises two
sub-populations (or {\em modules}) 
with the nature of interactions between spins
depending on whether they belong to the same or different
groups. Our central result is that when subjected to a uniform
magnetic field at a finite temperature, one of the sub-populations can 
become {\em ordered} while the other remains {\em disordered}.
This is surprising as both the interactions as well as the external
field for the two
modules are {\em identical}. Moreover, 
the chimera state is not a metastable state,
but rather the global minimum of free energy for the system.
The critical behavior of the system associated with the onset of
chimera ordering is established in this paper by an exact analytical treatment.
We also numerically demonstrate the existence of similar complex
ordering phenomena in three-dimensional spin systems with nearest neighbor
interactions.
This opens up the possibility of experimentally observing chimera
states in layered magnetic systems, e.g.,
manganites~\cite{kincaid75,stryjewski77}.

\begin{figure}
\begin{center}
\includegraphics[width=0.9\linewidth]{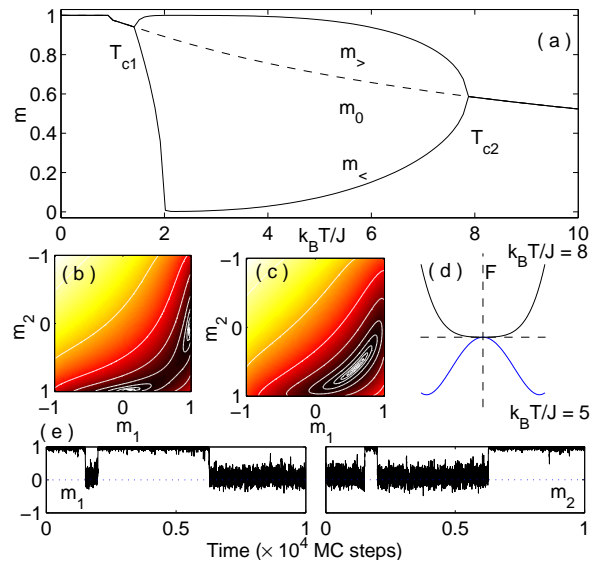}
 \end{center}
\caption{
(a) Variation of magnetization per spin of the two modules ($m_1$, $m_2$)
with temperature. The free energy landscape corresponding to chimera
order at $k_B T/J = 5$ (b) shows that there are two energy minima for
$m_1 \neq m_2$ (the curves are iso-energy contours and darker shades
correspond to lower energy), whereas outside the range
$[T_{c1},T_{c2}]$ there is
only one energy minima ($m_0$) on the $m_1 = m_2$ line as is seen for $k_B T/J
= 8$ (c). This is seen explicitly in (d) when the free energy per spin 
$F$ is observed
along the curve of steepest descent from $m_0$ (for $T_{c1} < T <
T_{c2}$) or along the curve of slowest ascent (for $T<T_{c1}$ or
$T>T_{c2}$). (e) In the chimera ordered state, the system switches between
the two energy minima corresponding to the two modules exchanging
their magnetization states between $m_{>}$ and $m_{<}$ shown for MC
simulations with $N=100$.
In all cases, $a = 1$ and $b = H = 10$.
}
\label{bifurcation}
\end{figure}

We consider a system of $2 N$ globally coupled Ising spins arranged 
into two sub-populations, each having $N$ spins, at a constant temperature $T$
and
subjected to a uniform external magnetic field
$H (>0)$. A dynamical system analogous to our model
has recently been analyzed by Abrams {\em et al}~\cite{Strog08}
where two clusters of identical oscillators, each maintaining a fixed
phase difference with the others, was shown to possess a chimera
state.
The interaction between two spins belonging to the same module
is ferromagnetic, having strength $J$ ($>0$), while that between
spins belonging to different modules is antiferromagnetic with
strength $-J^{\prime}$ ($<0$). It is obvious that in the absence of an external
field, the modules will be completely ordered in
opposite orientations at zero temperature.
As temperature is increased, the magnitude of the magnetizations for
the two modules will decrease by the same amount, eventually becoming
zero at a critical temperature, $T_c$.
In the presence of an external field $H$ that favors spins with $+$ve
orientation,
the module having negative magnetization will be
subjected to competition between (i) the field $H$ which attempts to 
align the spins along the $+$ve direction and (ii) the
antiferromagnetic interaction $J^{\prime}$ which is trying to do the
opposite.
For a suitable choice of the parameters $J$, $J^{\prime}$ and a
strong field $H>H_0$
(where $H_0$ is a threshold field), as
the temperature is increased from zero, the spins in both
modules initially remain ordered and are oriented in the {\em same}
direction.
Beyond a certain critical temperature $T_{c1}$, we observe that one module 
becomes more disordered relative to the other.
As the temperature increases further beyond a second critical
temperature $T_{c2}$, the two modules
again attain the same magnetization, which 
decreases gradually with $T$ [Fig.~\ref{bifurcation}~(a-d)].
The phase transitions at $T_{c1}$ and $T_{c2}$ are continuous,  
characterized by critical exponents $\alpha$ and $\beta$ which can be
derived exactly. 
For $H<H_0$, the spins in the two modules are oriented at $T=0$ in
opposite directions, although having the same magnitude. At any finite
temperature below $T_{c2}$, the module whose spins were initially oriented
opposite to the direction of the field is seen to be more disordered
than the other module. The same critical exponents as in the case of
$H>H_0$ are observed for the transition at $T_{c2}$, beyond which the
magnetization of the two
modules are same in magnitude and orientation.

For the system described above, the energy for a given configuration
of spins is
\begin{equation} 
E = - J \sum_{\stackrel{i,j,\alpha}{i\ne j}} S_{i\alpha} S_{j\alpha} + J^{\prime} \sum_{\stackrel{i,j,\alpha,\beta}{ \alpha \ne \beta}}  S_{i\alpha} S_{j\beta} 
-H  \sum_{i,\alpha} S_{i\alpha}, \label{Edef}
\end{equation}
where $i,j=1, 2, \cdots N$ labels the spins in a particular module and
$\alpha, \beta = 1,2$ identifies the two modules.
Since each spin is connected to every other spin, mean-field treatment
is exact for our effectively infinite-dimensional system.
Thus, the total free energy of the system can be expressed as:
\begin{eqnarray} 
F(m_1, m_2)  = - a N (m_1^2 + m_2^2) + b N m_1 m_2 -  \nonumber \\ 
      H N (m_1 + m_2) + N k_B T \left[ S(m_1) + S(m_2)\right],  \label{Fdef}
\end{eqnarray}
where $m_1$ and $m_2$ are the magnetizations (per spin) of the modules
1 and 2, $S(m) = \frac{1}{2} [ (1+m)\log(1+m) + (1-m)\log(1-m) ] -
\log 2$ is the entropy term, and $a = J (N-1)/2$,
$b = J^{\prime} N$ are system parameters ($k_B$ being the Boltzmann
constant).

To find the condition for equilibrium at a temperature $T$, the free
energy can be minimized with respect to $m_1$ and $m_2$ to obtain:
\begin{eqnarray}
- 2am_1 + b m_2 - H  + \frac{k_B T}{2} \log \frac{1+m_1}{1-m_1} = 0,
\label{min1}\\
- 2am_2 + b m_1 - H  +  \frac{k_B T}{2} \log \frac{1+m_2}{1-m_2} =  0. \label{min2}
\end{eqnarray}
Solving Eq.~\ref{min1} in terms of Eq.~\ref{min2}, we obtain the
one-dimensional map,
\begin{equation}  g(x) = \frac{1}{b} \left[ 2ax + H - \frac{k_B
T}{2}\log\frac{1+x}{1-x}\right],    \label{map1} \end{equation}
whose solutions of the form $g^2 (x) \equiv g(g(x)) = x$ 
give the extrema $m^*_1$ and $m^*_2$ of the free-energy $F$
(Eq.~\ref{Fdef}).
Numerical solution for the extrema values shows that for suitable
parameter values and $H>H_0$, the system has two critical temperatures
$T_{c1}$ and $T_{c2}$. For temperatures lower than $T_{c1}$ and 
above $T_{c2}$
the only fixed-point of the map $g^2$
is the unstable fixed point, $g(x) = x$, of Eq.~(\ref{map1}).
Thus, this solution corresponds to $m_1 = m_2 = m_0$ (say), where the
free energy $F(m_1, m_2)$ has a minimum. The value for $m_0$ is
obtained from 
\begin{equation} 
(-2a + b) m_0 - H + \frac{k_B T}{2} \log \frac{1+m_0}{1-m_0} = 0.
\label{min0} 
\end{equation}
However, in the temperature range $T_{c1} < T < T_{c2}$,
there are {\em two} types of solutions of $g^2$: (i) a stable fixed point
$m_1 = m_2 = m_0$ [obtained from Eq.~(\ref{min0})] corresponding to a
saddle point of the free energy function, and (ii) the pair of
unstable fixed points $m_1 \ne m_2$ which form a period-2 orbit of
Eq.~(\ref{map1}) corresponding to a
minimum of the free energy $F$. As one of ($m_1$ , $m_2$) is higher
and the other low, we obtain a chimera state where one module is
disordered ($m_<$) relative to the other module ($m_>$).
As shown in Fig.~\ref{bifurcation}~(a), the chimera
state occurs through subcritical pitchfork bifurcations
as the temperature is increased above $T_{c1}$ or decreased below 
$T_{c2}$.
For $H<H_0$, the system exhibits chimera ordering for $T>0$ and it 
has a single critical temperature at $T_{c2}$ above which the
magnetizations of the two modules become same.

By observing the free-energy $F(m_1,m_2)$ landscape in the range
$0\leq m_1,m_2\leq 1$, we obtain a clear physical picture of the 
transition to chimera ordering. 
The homogeneous state $m_1 = m_2 = m_0$ is a local extremum (i.e.,
$\partial F/\partial m_{1,2} = 0$) for the range of parameters
considered here. However, its nature changes from an energy minimum to a
saddle point as the temperature is increased beyond $T_{c1}$ and again
changes back to a minimum when temperature exceeds $T_{c2}$.
This is seen by looking at the matrix of the second derivatives of
free energy per site with respect to $m_1, m_2$:
\begin{equation}
{\cal H}\big|_{m_0} = \begin{pmatrix} 
A & b\\
b & A
\end{pmatrix},
\end{equation}
where $A = -2a + k_B T \frac{1}{1-m_0^2}$. 
The eigenvalues of this matrix are $\lambda_+ = A+b$
along the $m_1 = m_2$ line and $\lambda_- = A-b$ in the direction perpendicular
to it (parallel to $m_1 = - m_2$ line). 
Below $T_{c1}$ and above $T_{c2}$ both
eigenvalues are positive indicating that $m_0$ is a minimum.
The transition to chimera ordering occurs in the range $T_{c1} < T <
T_{c2}$ when the smaller eigenvalue
$\lambda_-$ becomes negative while the other eigenvalue remains
positive, indicating that $m_0$ is now a saddle point. 
This gives us an implicit relation for $T_c$ as the temperature where
$\lambda_- = 0$, which gives 
$$k_B T_c = 2(2a+b)(1-m_0^2).$$
Numerical
investigation of the landscape indicates that this transition is
accompanied by
the creation of two minima away from the $m_1 = m_2$
line [Fig.~\ref{bifurcation}~(b-d)]. These minima are symmetrically
placed about the $m_1 = m_2$ line [as $F(m_1, m_2) = F(m_2, m_1)$] and
correspond to the two coexisting chimera
states $C_1: m_1 = m_>, m_2 = m_<$ and $C_2: m_1 = m_<, m_2 = m_>$.
The two minima are separated by an energy barrier
$\Delta = F(m_0,m_0)-F(m_>,m_<)$ which for a finite system can be
crossed by thermal energy. This switching behavior between the two
chimera states has a
characteristic time $\tau \sim {\rm exp} (\Delta/k_B T)$ which is
indeed observed from Monte Carlo (MC) simulations
[Fig.~\ref{bifurcation}~(e)].
Note that, each minima corresponds to a state with coexisting order and
disorder, and hence is unlike the minima seen in phase-coexistence
state of systems such as metamagnets, where each of the minima
represent a certain type of order~\cite{note1}.

\begin{figure}
\begin{center}
\includegraphics[width=0.99\linewidth]{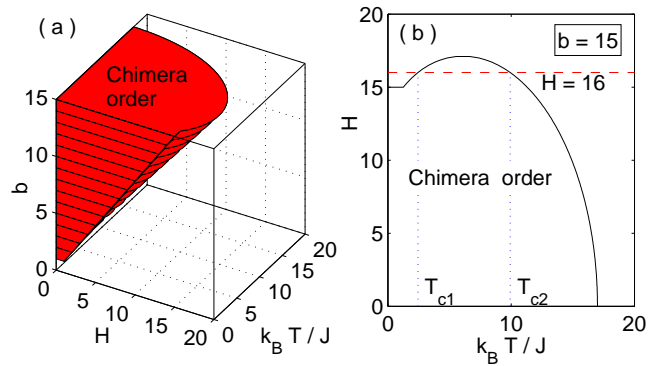}
 \end{center}
 \caption{(a) Phase diagram in the magnetic field ($H$), temperature ($k_B
 T / J$) and anti-ferromagnetic coupling ($b$) parameter space 
 obtained by numerical minimisation of free energy for $a = 1$
 with the region in which chimera ordering occurs being indicated.
 A cross-section along $H-T$ plane for $b=15$ is shown in (b). The broken
 line indicates $H = 16$, for which the critical temperatures are
 shown by dotted lines.}
 \label{phase_diag}
\end{figure}
Fig.~\ref{phase_diag} shows the region in $(H-T-b)$ parameter space
where chimera ordering is observed in our system as obtained by
numerical minimization of the free energy.
Temperature induced transitions are always continuous whose
exponents are analytically derived below.
To investigate the critical behavior of the system around $T_{c1}$ and
$T_{c2}$, we shall use the order parameters:
$$p_1 = m_1 - m_2 \ \ \ {\rm and}\ \ \ p_2 = 2 m_0 - (m_1 + m_2).$$ 
For $T_{c1} < T < T_{c2}$ where the chimera ordering is observed, as
mentioned earlier the free energy minima are at $m_1$ and
$m_2$ while $m_0$ corresponds to a saddle point. The order parameters 
$p_1$ and $p_2$ are non-zero in this region and 
and zero elsewhere. 
When $p_1, p_2$ are small, we solve for them using
Eqs. (\ref{min1}), (\ref{min2}) and (\ref{min0}) by 
expressing $m_1$ and $m_2$ in terms of $p_1, p_2$, and obtain
\be p_1 \propto \mid T - T_c \mid ^{1/2} \;\;\;\; {\rm and } \;\;\;\;
\mid p_2\mid \propto \mid T - T_c \mid.  \ee
Thus, as $T \rightarrow T_{c1}^+$ or $T \rightarrow T_{c2}^-$,
the order parameters vanish continuously with exponents $\beta = 1/2$
for $p_1$ and $\beta = 1$ for $p_2$.
Similar calculations for the field induced transition at finite
temperature yield identical critical exponents. Note that,
at zero temperature the field induced transition is of first order
and its discontinuous nature
can be shown exactly by analyzing the free energy.
The values of the exponents for all continuous transitions have been
confirmed by us numerically.

We have also analyzed the critical behavior of the specific heat $C =
-T \frac{\partial^2 F_0}{\partial T^2}$ where
$F_0$ is the equilibrium free energy at a given $a$, $b$, $H$ and $T$.
Although it involves both first and second order derivatives of $p_1$
and $p_2$, as the most dominant term is $\partial^2 p_1/\partial T^2$, the
divergence at critical temperature is characterized by exponent
$\alpha=3/2 : C \propto \mid T - T_{c1,c2} \mid ^{-3/2}$. 

\begin{figure}
\begin{center}
\includegraphics[width=0.99\linewidth]{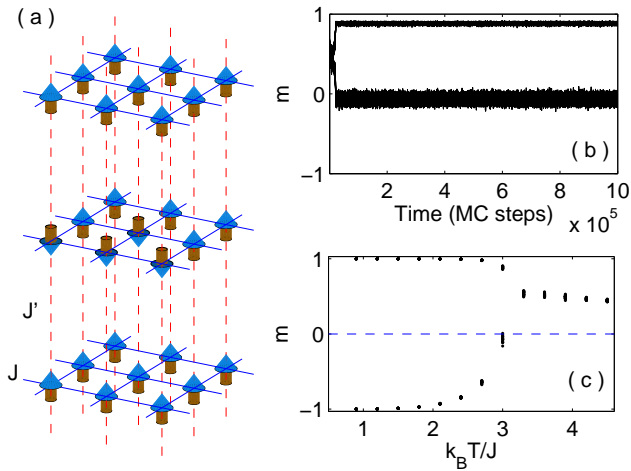}
  \end{center}
\caption{(a) Schematic diagram of the 3-dimensional layered spin
system with ferromagnetic (anti-ferromagnetic) interactions within
layers, $J$ (between layers, $J^{\prime}$) indicated by continuous
(broken) lines. In the chimera state alternate layers show order and disorder.
(b) The time-evolution in MC steps of the different layers of a system
with 32 planes each having $128 \times 128$ spins showing chimera
ordering for $k_B T /J = 3$. (c) The magnetizations of different layers
of the $128 \times 128 \times 32$ spin system at different
temperatures. Chimera ordering is manifested as different values of
$|m|$ for alternate layers (e.g., at $k_B T/J = 3$).
In all cases $J=J^{\prime}=1$ and $H=1.8$.
}
\label{Fig3d}
\end{figure}

While the system we have considered so far has the advantage of being
amenable to exact analytical treatment, we have also numerically
analyzed spin models which are closer to real magnetic materials.
We have performed MC simulation studies of a
three-dimensional Ising spin model with nearest neighbor interactions
having an anisotropic nature. The system emulates a layered magnetic
system comprising multiple layers of two-dimensional spin arrays
stacked on top of each other, 
with interactions along a plane being ferromagnetic ($J$) and
those between planes anti-ferromagnetic ($-J^{\prime}$).
Fig.~\ref{Fig3d} shows chimera ordering in such a 3-dimensional spin system
with periodic boundary conditions.
Similar behavior was observed in other systems having different sizes,
parameters and interaction structure, indicating that chimera ordering
is a robust phenomenon that should be possible to observe in an
experimental magnetic system.

In summary, we have shown the existence of a novel complex
ordering behavior that we term chimera order in analogy with the
coexistence of coherent and incoherent behavior in dynamical systems.
For a system of two clusters of Ising spins, where the spins are
coupled ferromagnetically (anti-ferromagnetically) to all spins in the
same (other) cluster, subjected to a uniform external magnetic field at a given
temperature, chimera ordering is manifested as a much higher magnetization
in one cluster compared to the other. To illustrate the wider
implication of our result we can use the analogy of two communities of
individuals who are deciding between a pair of competing choices. 
The interactions of an agent with other members of its own community 
strongly favor consensus while that with members of a different
community are antagonistic. Thus, given that every individual is
exposed to the same information or external environment,
we would expect that unanimity about a particular choice in one community
will imply the same for the contrary choice in the other community.
However, the occurrence of chimera order suggests that under certain
conditions, when given the same external stimulus
we may observe consensus in one community while the other
is fragmented.
The generality of the chimera state as defined in our paper suggests
that it may be experimentally observed in physical systems. Our
demonstration of chimera order in a three-dimensional spin system with 
nearest neighbor interactions indicate that a possible experimental
example can be layered magnetic materials (e.g., manganites) 
having different types of
interactions between and within layers~\cite{kincaid75,stryjewski77}.
Although the present paper looks at the case of two competing choices,
it is possible to extend the analysis to $q$-state Potts spin
dynamics. Given the wider applicability of spin models for studying
ordering in different contexts, one can consider other connection topologies
as well as mesoscopic features such as the occurrence of multiple
modules ($>2$) and hierarchical organization.

We thank I. Bose, G. Menon and P. Ray for helpful discussions.
This work was supported in part by CSIR, UGC-UPE, IMSc Associate
Program and IMSc Complex Systems (XI Plan) Project.



\end{document}